\documentclass[%
 reprint,superscriptaddress,%groupedaddress,
 amsmath,amssymb,prx,longbibliography]{revtex4-1}

\usepackage{physics}
\usepackage{xcolor}
\usepackage{bm}
\usepackage{xtab,afterpage,longtable}
\usepackage{lipsum}
\usepackage{graphicx}% Include figure files
\usepackage{dcolumn}% Align table columns on decimal point
\usepackage{booktabs} % For professional looking tables
\usepackage{multirow}
\usepackage{color}
\usepackage{eucal}
\usepackage{mathrsfs}
\usepackage{siunitx}
\usepackage{makecell}
\usepackage{hyperref}% add hypertext capabilities
\renewcommand{\ap}{a}
\begin{document}

\title{Non-Hermitian Spin-Spin Interaction Mediated by Chiral Phonons}% Force line breaks with \\

\author{Haowei Xu}%
\affiliation{Department of Nuclear Science and Engineering, Massachusetts Institute of Technology, Cambridge, MA 02139, USA}

\author{Guoqing Wang}%
\affiliation{Department of Nuclear Science and Engineering, Massachusetts Institute of Technology, Cambridge, MA 02139, USA}
\affiliation{Research Laboratory of Electronics, Massachusetts Institute of Technology, Cambridge, Massachusetts 02139, USA}
\affiliation{Department of Physics, Massachusetts Institute of Technology, Cambridge, Massachusetts 02139, USA}

\author{Changhao Li}
\affiliation{Department of Nuclear Science and Engineering, Massachusetts Institute of Technology, Cambridge, MA 02139, USA}
\affiliation{Research Laboratory of Electronics, Massachusetts Institute of Technology, Cambridge, Massachusetts 02139, USA}

\author{Hao Tang}%
 \affiliation{Department of Materials Science and Engineering, Massachusetts Institute of Technology, Cambridge, MA 02139, USA}

\author{Paola Cappellaro}
\email{pcappell@mit.edu }
\affiliation{Department of Nuclear Science and Engineering, Massachusetts Institute of Technology, Cambridge, MA 02139, USA}
\affiliation{Research Laboratory of Electronics, Massachusetts Institute of Technology, Cambridge, Massachusetts 02139, USA}
\affiliation{Department of Physics, Massachusetts Institute of Technology, Cambridge, Massachusetts 02139, USA}

\author{Ju Li}%
\email{liju@mit.edu}
\affiliation{Department of Nuclear Science and Engineering, Massachusetts Institute of Technology, Cambridge, MA 02139, USA}
\affiliation{Department of Materials Science and Engineering, Massachusetts Institute of Technology, Cambridge, MA 02139, USA}

\date{\today}% It is always \today, today,

\begin{abstract}
Non-Hermiticity and chirality are two fundamental properties known to give rise to various intriguing phenomena. However, the interplay between these properties has been rarely explored. In this work, we bridge this gap by introducing an off-diagonal non-Hermitian spin-spin interaction mediated by chiral phonons. This interaction arises from the spin-selectivity due to the locking between phonon momentum and angular momentum in chiral materials. The resulting non-Hermitian interaction mediated by the vacuum field of chiral phonons can reach the kHz range for electron spins and can be further enhanced by externally driven mechanical waves, potentially leading to observable effects in the quantum regime. Moreover, the long-range nature of phonon-mediated interactions enables the realization of the long-desired non-Hermitian interaction among \emph{multiple} spins. The effect proposed in this work may have wide-ranging applications in cascaded quantum systems, non-Hermitian many-body physics, and non-Hermitian cooling.
\end{abstract}

\maketitle

\textbf{Introduction.} Recently, non-Hermitian physics in open quantum systems has attracted significant interest. Well-controlled non-Hermiticity, such as balanced loss and gain, can engender a wealth of intriguing phenomena, such as exceptional points~\cite{miri2019exceptional,chen2017exceptional,xu2016topological}, non-Hermitian skin effect~\cite{zhang2022review,xiao2020non,weidemann2020topological}, non-Hermitian topology~\cite{bergholtz2021exceptional,shen2018topological,kawabata2019symmetry}, and non-Hermitian cooling~\cite{xu2024exponentially}. However, quantum engineering of non-Hermitian interactions is a highly non-trivial task, especially for off-diagonal  interactions among multiple quantum objects. In practice, such interactions can be realized through reservoir engineering~\cite{metelmann2015nonreciprocal} or parametric driving~\cite{kamal2011noiseless,sliwa2015reconfigurable}, but are predominantly limited to systems with no more than two quantum objects.

Chirality, the asymmetry between an object with its mirror image, has significant implications across various disciplines. A well-known example is the chiral-induced spin selectivity~\cite{ray1999asymmetric,naaman2019chiral,liu2021chirality}.  Another key characteristic of solid-state chiral materials is the presence of \emph{truly} chiral phonons~\cite{zhang2014angular,zhang2015chiral,ishito2023truly,ueda2023chiral}. The coupling of chiral phonons to itinerant electron spins and the subsequent impact on transport properties has been studied by several groups ~\cite{kim2023chiral,nomura2023nonreciprocal}. However, the correlation between chiral phonons and \emph{localized} electron and/or nuclear spins, which are crucial qubit platforms for quantum engineering ~\cite{wolfowicz2021quantum,chatterjee2021semiconductor}, remains largely unexplored.

\begin{figure}
    \centering
    \includegraphics[width=0.7\linewidth]{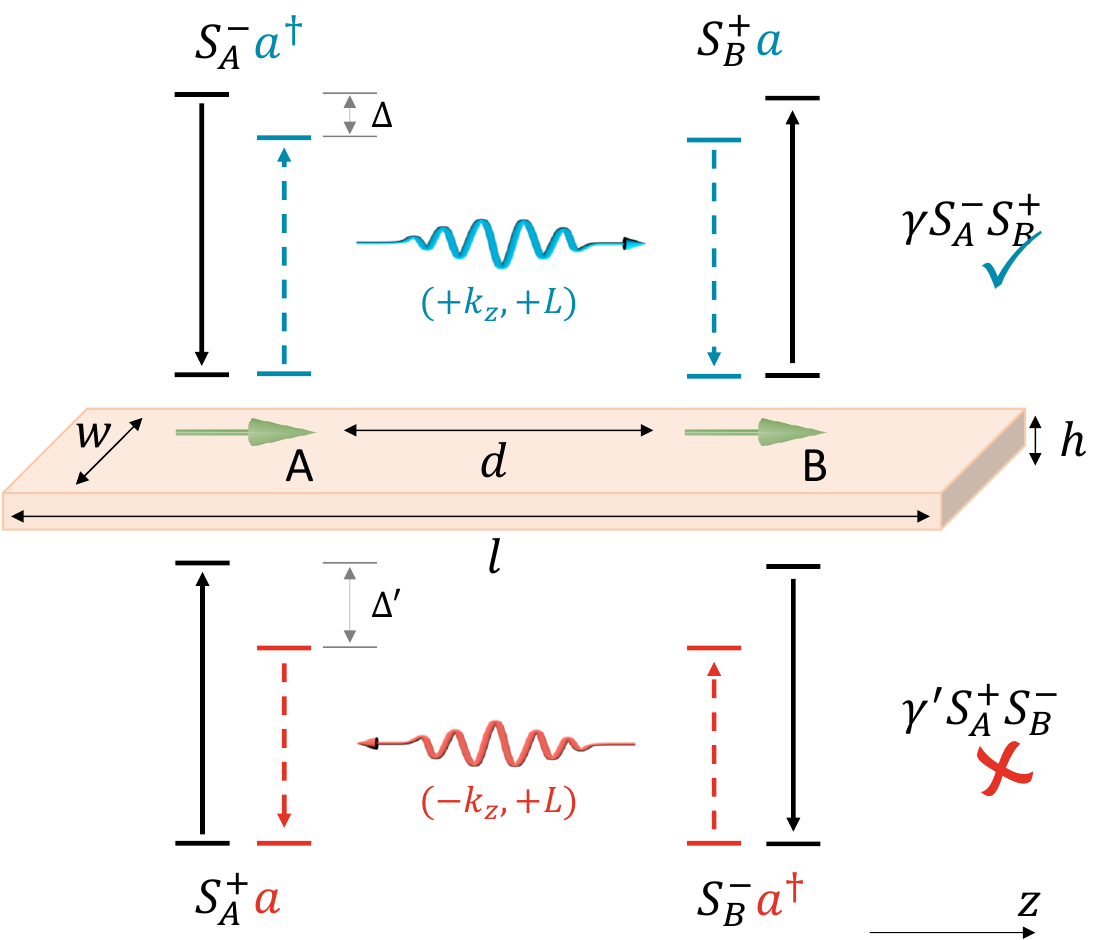}
    \caption{Non-Hermitian spin-spin interactions mediated by chiral phonons. Two spins A and B are embedded in a chiral material with a distance of $d$. The right-propagating $(+k_z, +L)$ phonon mode (colored blue) leads to an effective interaction $S_A^{-}S_B^{+}$, which results from spin-phonon interactions $S_A^{-}a^{\dagger}$ and $S_B^{+}a$.  The left-propagating $(-k_z, +L)$ phonon (colored red) would have induced the reverse $S_A^{+}S_B^{-}$ interaction, but the strength can be rather small due to the large detuning $\Delta'$. Consequently, the interaction between spins A and B is non-reciprocal. Black arrows denote spin transitions.}
    \label{fig:non-reciproal}
\end{figure}

In this work, we propose a non-Hermitian spin-spin interaction mediated by chiral phonons. The angular momentum of chiral phonons results in selective interactions with spins. Intuitively, for two spins A and B separated by a distance, a right-propagating phonon with positive angular momentum $L$ can transfer angular momentum from spin A to spin B, but not in the reverse direction. This creates an effective non-reciprocal spin-spin interaction  (Figure~\ref{fig:non-reciproal}). In non-chiral materials, reciprocity is restored when considering a left-propagating phonon with the same $L$, which transfers the angular momentum from spin B to A. However, in chiral materials, right- and left-propagating phonons exhibit distinct properties. We will demonstrate that the strength of spin-spin interactions mediated by them can differ by several orders of magnitude, resulting in a highly non-Hermitian total interaction (Table~\ref{tab:four_modes}). Notably, the phonon-mediated interaction is long-range, enabling the realization of non-Hermitian interactions for multiple spins coupled to a common phonon mode.

In the following sections, we will first introduce the unique properties of chiral phonons in typical chiral materials, such as $\alpha$-$\rm SiO_2$. Next, we will demonstrate the phonon-mediated interactions for both electron spins and nuclear spins. Particularly, we will emphasize how these interactions can become non-Hermitian due to the locking of phonon momentum and angular momentum in chiral materials. Finally, we will discuss the experimental feasibility of such a non-Hermitian spin-spin interaction. We find that the strength of the non-Hermitian interaction can reach $~\rm kHz$ level for electron spins embedded in mechanical resonator with micro-meter dimensions, which can result in observable effects considering that the electron spin decoherence time can reach milliseconds at cryogenic temperature. For nuclear spins, the interaction mediated by the zero-point field of the resonator is relatively weak, and we propose to enhance the interaction by externally pumping the mechanical waves. We will also explore potential applications of the effect revealed in this work, including its implications for non-Hermitian many-body physics.

\begin{figure}
    \centering
    \includegraphics[width=0.8\linewidth]{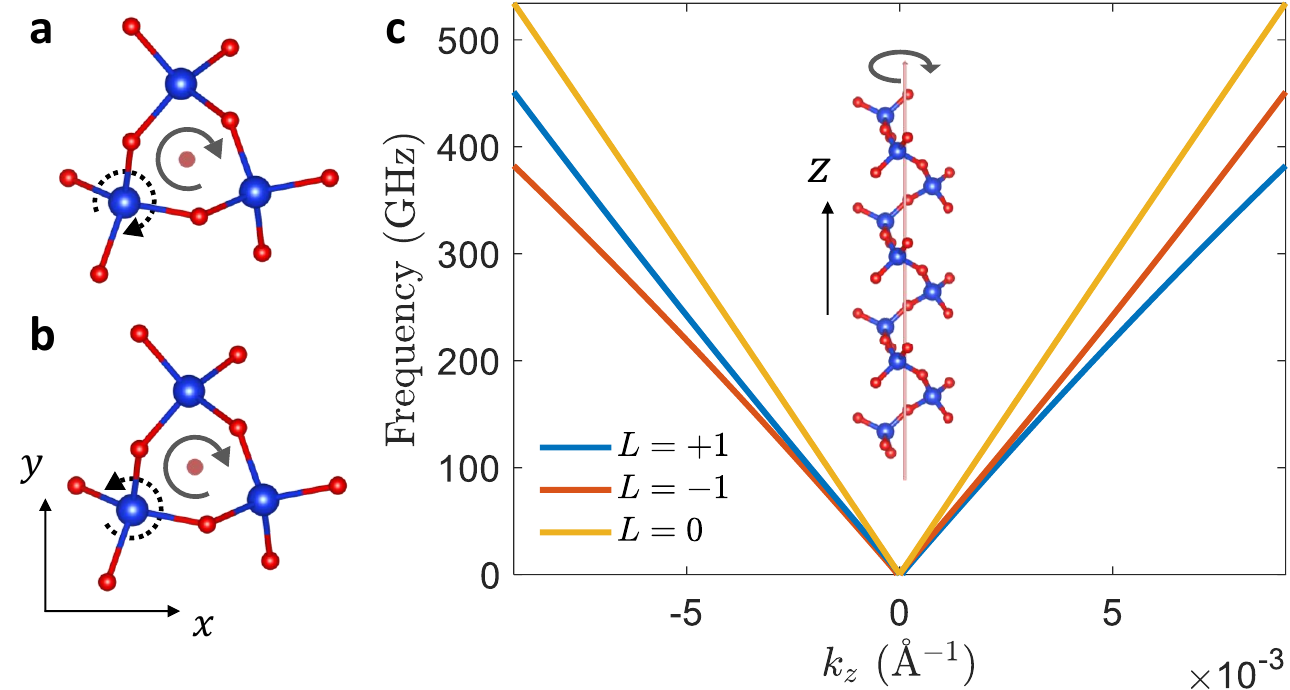}
    \caption{Chiral phonons in chiral materials. (a, b) Top view of the atomic structure of $\alpha$-$\rm SiO_2$. The grey solid arrow indicates the spiral of the -Si-O-Si-chain, whose side view is  shown in the inset of (c). The black dashed arrows indicate the circular atomic vibration of phonons with (a) $L=+1$ and (b) $L=-1$. The atomic vibration either co-rotate or counter-rotate with the spiral static atomic structure, lead to different phonon frequencies. (c) Acoustic phonon dispersion of  $\alpha$-$\rm SiO_2$. The $L=0$ modes is longitudinal, while the two transverse modes have $L=\pm 1$.  }
    \label{fig:chiral_phonon}
\end{figure}

\textbf{Chiral Phonons.} Like photons, phonons can carry angular momentum. If the atomic vibration along $x$ and $y$ directions have a $\pm \pi/2$ phase difference, then the rotational motion will lead to a non-zero angular momentum (Figure~\ref{fig:chiral_phonon}a, b). Phonons with clockwise and counter-clockwise atomic vibration would have negative and positive angular momentum, respectively. This is in analogy to the helicity of circularly polarized light. In crystals, there are two definitions of phonon angular momentum. The genuine angular momentum is defined by $\sum_i m_i \bm{r}_i \times \bm{v}_i$ , where $m_i$, $\bm{r}_i$, and $\bm{v}_i$ are the mass, displacement and velocity of the $i$-th atom, respectively~\cite{zhang2014angular}. Meanwhile, at high-symmetry points of the Brillouin zone, the phonon eigenmodes are eigenvectors of rotational symmetry operations, and the pseudoangular momentum (PAM) $L$ is defined by the corresponding eigenvalue~\cite{zhang2015chiral}, which is typically an integer $L=0, \pm 1$. The angular momentum selection rules depends on PAM~\cite{streib2021difference,ishito2023truly}, which will be the focus hereafter.

In chiral materials, the atomic structure cannot be superimposed onto its mirror image. A typical chiral material is $\alpha$-quartz ($\alpha$-$\rm SiO_2$), where the -Si-O-Si- chains spiral along the $c$-axis ($z$-direction) with either clockwise or counterclockwise chirality (Figure~\ref{fig:chiral_phonon}). This structural chirality would correlate with the rotational atomic motions of the chiral phonons. Depending on the PAM $L$, a phonon propagating along the $z$-direction the would experience different restoring forces and consequently have different frequencies.
%These phonons are truly chiral, as the they are converted to their enantiomers by spatial inversion, instead of time-reversal combined with rotations~\cite{barron2009molecular,ishito2023truly}.
These arguments are verified by our \emph{ab initio} calculations. For phonons propagating along the $z$-direction, the longitudinal acoustic mode $L=0$ has the highest frequency with a sound speed of around $5.8\times 10^3~\rm m/s$, consistent with experimental results. On the other hand, the two transverse acoustic (TA, $L=\pm 1$) modes are not degenerate in frequency, and their group velocities are different by as large as $20~\%$.  This is in sharp contrast to conventional non-chiral materials such as diamond and cubic Silicon, where the two TA modes are degenerate in frequency. The group velocities of TA modes of several typical chiral materials, including $\alpha$-$\rm HgS$ and $\alpha$-$\rm TeO_2$, are summarized in the last three columns of Table~\ref{tab:four_modes}.

More importantly, for $+k_z$ phonons with positive momentum and velocity, the $+L$ TA mode has lower frequency, whereas for $-k_z$ phonons with negative momentum, the $-L$  TA mode has the lower frequency. This is enforced by time-reversal symmetry, which maps a $(+k_z, +L)$ phonon to a $(-k_z, -L)$ phonon. The locking between phonon momentum and angular momentum, which can be represented by $k \cdot L$, is reminiscent of the spin-orbit interaction observed in photons~\cite{bliokh2015spin,lodahl2017chiral}.
%However, unlike photons, where the spin-orbit interaction often requires sophisticated nanostructures,
Note the $k\cdot L$ locking of phonons is an intrinsic property of chiral materials and does not require sophisticated nanostructures. It can lead to intriguing phenomena, as we will explain below.

%For non-chiral phonons, the atomic vibrations are linear, leading to vanishing angular momentum. For chiral phonons, the  between $x$ and $y$  directions results in rotational motion and non-zero phonon angular momentum.

%breaks the degeneracy in phonon modes with positive and negative angular momenta. Hence, the phonon modes have well-defined angular momenta.  (Figure X). The computed phonon acoustic phonon dispersion of  $\alpha$-$\rm SiO_2$ is plotted in Figure X, where both geniune and pesudo-angular momentum

\textbf{Chiral Spin-Phonon Interactions}. Next, we consider the mechanisms for spin-phonon coupling. We assume that the electron or nuclear spins are hosted by point defects, and that the point defects do not affect the phonon dynamics of the host material. This is particularly true for long wavelength acoustic phonons, which will be the focus of this work. Because of the circular motion of the charge, chiral phonons carry intrinsic magnetic moments~\cite{juraschek2019orbital,ren2021phonon,cheng2020large,baydin2022magnetic}, which can directly interact with spins. However, as we show in the Supplementary Information (SI) Section A, the magnetic dipole interaction between spins and phonons are rather weak.

\begin{figure}
    \centering
    \includegraphics[width=1\linewidth]{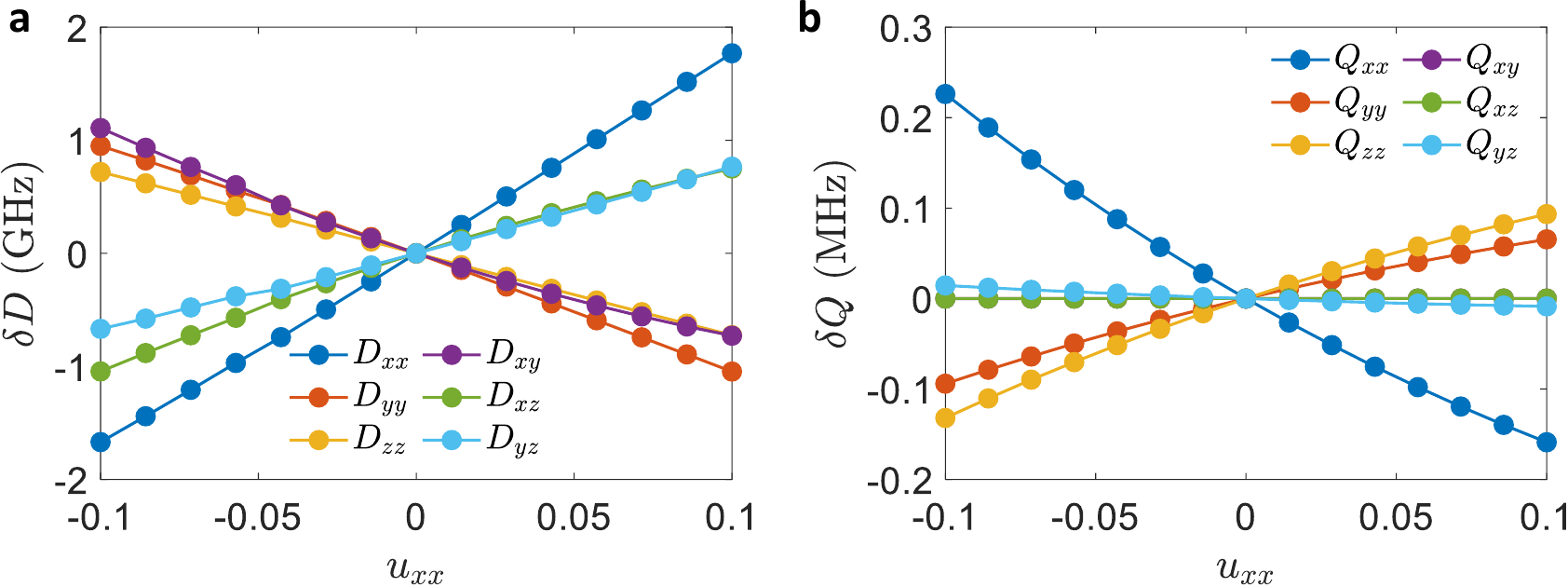}
    \caption{(a) Zero field splitting tensor $D$ of $\rm C^0_O$ in $\alpha$-$\rm SiO_2$ as a function of strain $u_{xx}$. Nuclear quadrupole tensor $Q$ of $\rm Al_{Si}^{-}$ in $\alpha$-$\rm SiO_2$ as a function of strain $u_{xx}$. The curves are offset to the value at $u_{xx}=0$.}
    \label{fig:strain-dependence}
\end{figure}

For electron spins, a better strategy is to use the zero-field splitting $H_e = \sum_{ij} D_{ij} S_i S_j$~\cite{wertz2012electron, bennett2013phonon,kepesidis2013phonon,maity2020coherent,maity2022mechanical}, where the subscripts are Cartesian indices.  Phonon, as strain and/or lattice distortion, can induce variances in the $D$ tensor, which then influences the electron spin. This is described by $H_{Sp} = \sum_{ij,kl} \Xi^{S}_{ij,kl} S_i S_j u_{kl}$, where $ \Xi^{S}_{ij,kl} \equiv \frac{\partial D_{ij}}{\partial u_{kl}}$ is the response function and $u_{kl}$ is the strain tensor.  As a example, we examine the charge neutral Carbon substitution for Oxygen ($\rm C^0_O$) in $\alpha$-$\rm SiO_2$, which has an electron spin $S=1$ according to our simulations. In Figure~\ref{fig:strain-dependence}a, different components of the $D$ tensor are plotted as a function of the uni-axial strain $u_{xx}$. One can see the response function $\Xi^S$ is on the order of 10~GHz, comparable with that of Nitrogen-Vacancy centers in diamond~\cite{bennett2013phonon}. For nuclear spins, the nuclear quadrupole interaction~\cite{krane1991introductory} is also strain dependent, leading to the nuclear spin-photon interaction $H_{Ip} = \sum_{ij,kl} \Xi^{I}_{ij,kl} I_i I_j u_{kl}$, where $ \Xi^{I}_{ij,kl} \equiv \frac{\partial Q_{ij}}{\partial u_{kl}}$ is the response function and $Q_{ij}$ is the nuclear quadrupole tensor. We consider $\rm Al^{-}_{Si}$ in $\alpha$-$\rm SiO_2$, i.e., the Aluminum substitution for Silicon with a negative charge. $\rm Al^{-}_{Si}$ has electron spin $S=0$, so that unwanted hyper-fine interactions can be avoided. Meanwhile, $\rm ^{27}Al$ has nuclear spin of $I=5/2$ and a relatively large nuclear quadrupole momentum of $q\approx 146~\rm mb$. The nuclear quadrupole tensor $Q$ of $\rm ^{27}Al$ is plotted in Figure~\ref{fig:strain-dependence}b as a function of $u_{xx}$, where one can see the response function $\Xi^I$ is on the order of 1 MHz. Note that the magnitude of the response functions will depend on specific defect systems. In the following, we will adopt for an order-of-magnitude estimation
\begin{equation}\label{eq:response_functions}
\begin{aligned}
    \Xi^{S} \equiv \frac{\partial D}{\partial u} & = 10~\mathrm{GHz}, \\
    \Xi^{I} \equiv \frac{\partial Q}{\partial u} & = 1~\mathrm{MHz}. \\
\end{aligned}
\end{equation}
For brevity, we will occasionally use $\Xi$ to denote the response function and use $S$ to denote both electron and nuclear spins.

As discussed before, chiral acoustic phonons carry PAM $L=\pm 1$~\cite{zhang2014angular,zhang2015chiral}. Due to the angular momentum conservation law~\cite{streib2021difference,ishito2023truly}, $S$ must be lowered by 1 when a chiral phonon with $L=+1$ is created via spin-phonon interactions. Consequently, $H_{Sp}$ can be simplified as
\begin{equation}\label{eq:spin_phonon}
H_{Sp} = g (S^{-} \ap^{\dagger} + S^{+} \ap),
\end{equation}
where $S^{\pm}$ are spin raising and lowering operators, while $\ap^{\dagger}$ ($\ap$) is the creation (annihilation) operator of the $L=+1$ chiral phonon. Meanwhile, $g \equiv \Xi u^{\rm zpf}$ is the reduced spin-phonon interaction strength, where $u^{\rm zpf}$ is the zero-point strain field of the chiral phonon mode. For a $L=-1$ chiral phonon, the angular momentum conservation law enforces terms like $S^{+}a^{\dagger}$. However, these counter-rotating terms induce large energy detuning, that is, they strongly violate energy conservation.

\textbf{Coherent Phonon-Mediated Spin-Spin Interactions.} The spin-phonon interaction can induce effective spin-spin interaction~\cite{rabl2010quantum,kuzyk2018scaling,rosenfeld2021efficient,fung2023programmable}. To be specific, we consider two spins $A$ and $B$  interacting with a phonon with angular momentum $L=+1$ (Figure~\ref{fig:non-reciproal}). We assume the two spins have the same spin splitting. In the rotating frame of the phonon frequency, the total Hamiltonian of the spin-phonon system is
\begin{equation}\label{eq:spin_phonon_total_Hamitonian}
\begin{aligned}
  H = \Delta (S_A^{z} + S_B^{z}) + g (S^{-}_A \ap^{\dagger} + S^{+}_A \ap) + g (S^{-}_B \ap^{\dagger} + S^{+}_B \ap).
\end{aligned}
\end{equation}
Here $\Delta$ is the frequency detuning between the spins and the phonon. The subscripts denote spins A and B, and we assume the spin-phonon coupling strength is the same for spins A and B. In the large-detuning limit ($\Delta \gg g$), the effective spin-spin interaction can be expressed as~\cite{carmichael1993quantum,brion2007adiabatic}
\begin{equation}\label{eq:spin_spin}
\begin{aligned}
  H_{AB} = i\gamma  (S^{+}_A S^{-}_B - S^{-}_A S^{+}_B),
\end{aligned}
\end{equation}
where $\gamma = \frac{2g^2}{\Delta}$ is the effective coupling strength. This can be understood as follows: Spin A first (virtually)  gives its spin angular momentum to the phonon (via $g S^{-}_A \ap^{\dagger}$). Then, the phonon transfers the angular momentum to spin B and is annihilated (via $g S^{+}_B \ap$). The effective spin-spin interaction in Eq.~\ref{eq:spin_spin} is obtained when the phonon operators are adiabatically eliminated~\cite{winkler2003quasi}.  In principle, the phonon-mediated interaction can be realized as long as the two spins are coupled to a common phonon mode. This is reminiscent of the phonon-mediated interactions in trapped ions~\cite{cirac1995quantum,monroe2021programmable} and the cavity photon-mediated interactions among atoms~\cite{van2013photon,evans2018photon}.

\begin{table*}
    \centering
      \renewcommand{\arraystretch}{1.5}

\caption{Four different TA phonon modes and their contributions to the spin-spin interaction. The second column shows the spin-phonon interaction for spin A that are allowed by angular momentum conservation. The corresponding interaction for spin B can be obtained by a Hermitian conjugate and replacing subscript $A$ with $B$.
The frequency detuning and spin-spin interaction strength columns are for electron spins. In the last three columns, the  group velocities of each phonon mode in three typical chiral materials are listed in the unit of $\rm 10^3~m/s$.}
\label{tab:four_modes}
    \begin{tabular}{>{\centering\arraybackslash}p{2cm} |>  {\centering\arraybackslash}p{2cm}>{\centering\arraybackslash}p{2cm} | >{\centering\arraybackslash}p{2cm}>{\centering\arraybackslash}p{3cm}|>{\centering\arraybackslash}p{1.5cm}>{\centering\arraybackslash}p{1.5cm}>{\centering\arraybackslash}p{1.5cm}} \hline \hline
       Phonon mode&  Spin-phonon interaction   & Spin-spin interaction & Frequency detuning & Spin-spin interaction strength& $\alpha$-$\rm SiO_2$ & $\alpha$-$\rm HgS$ & $\alpha$-$\rm TeO_2$\\ \hline
         $(+k_z, +L)$&   $S^-_A a^{\dagger}$ & $S_A^{-} S_B^{+}$&$\sim 10~\rm kHz$ &  $\gamma\sim [0.1-1]~\rm kHz$&  4.2&  1.3& 2.5\\ \hline
         $(-k_z, +L)$&   $S_A^+ a$ & $S_A^{+} S_B^{-}$& $\sim 0.1~\rm GHz$ &   $\gamma' < 1~\rm Hz$&  5.0&  1.6& 2.4\\ \hline
         $(+k_z, -L)$&   $S_A^+ a^{\dagger}$ & $S_A^{+} S_B^{-}$& $\sim 10~\rm GHz$&  $\ll \gamma, \gamma'$&  5.0&  1.6& 2.4\\ \hline
         $(-k_z, -L)$&   $S_A^{-}a$ & $S_A^{-} S_B^{+}$& $\sim 10~\rm GHz$&  $\ll \gamma, \gamma'$&  4.2&  1.3& 2.5\\ \hline  \hline
    \end{tabular}

\end{table*}

\textbf{Non-Hermitian Spin-Spin Interactions}. So far, the spin-phonon and spin-spin interactions remain Hermitian. As indicated by Eq.~(\ref{eq:spin_spin}), if spin A can influence spin B, then spin B can influence spin A in the same manner. How can the phonon-mediated spin-spin interactions be non-Hermitian?

In this regard, we propose to use yet another degree-of-freedom of phonons, namely their momentum $k$, which determines the direction of phonon propagation. For definiteness, we assume spin A is located on the left hand side of spin B with a distance of $d$ (Figure~\ref{fig:non-reciproal}). Intuitively, a phonon traveling to the right ($+k_z$) would first interact with spin A, and then interact with spin B after a time delay of $d/v$ with $v$ the phonon velocity. Due to causality, one can argue that the right-propagating phonon transfers the influence only from spin A to B, but not from B to A, which would require a left-propagating phonon.

For a $(+ k_z, + L)$ phonon,
%where $+k_z$ ($-k_z$) labels right (left) propagation, while $+L$ ($-L$) labels positive (negative) phonon angular momentum. We first consider a $(+k, +L)$ phonon.
The phonon-induced directional spin-spin coupling is a cascaded interaction and can be described by the master equation~\cite{carmichael1993quantum,gardiner1993driving,stannigel2012driven,guimond2016chiral}
\begin{equation}\label{eq:master_equation}
\begin{aligned}
\frac{\partial \rho} {\partial t} = -i [H_{AB}, \rho] + 2\gamma\mathcal{D}[z] \rho,
\end{aligned}
\end{equation}
where $\rho$ is the density matrix of spins A and B, while $\mathcal{D}[o]\rho \equiv o\rho o^{\dagger} - \frac{1}{2} (o^{\dagger} o \rho + \rho o^{\dagger} o)$ is the Lindblad super-operator for an operator $o$. Meanwhile, one has
\begin{equation}\label{eq:HAB_sigma_with_phi}
\begin{aligned}
  H_{AB} & = i\gamma  ( e^{-ik_zd} S^{+}_A S^{-}_B - e^{ik_zd} S^{-}_A S^{+}_B), \\
  z & =   S^{-}_A  + e^{-ik_zd} S^{-}_B.
\end{aligned}
\end{equation}
Compared with Eq.~(\ref{eq:spin_spin}), here we introduced an additional factor  $e^{-ik_zd}$ to account for phase acquired by phonon propagation. Meanwhile, the jump operator $z$ describes the \emph{collective} decay of the spins due to the coupling with the common $(+ k_z, + L)$ phonon mode.  Eqs.~(\ref{eq:master_equation}, \ref{eq:HAB_sigma_with_phi}) combined describe a non-Hermitian interaction between spins A and B. This can be better demonstrated by re-writing the master equation as~\cite{metelmann2015nonreciprocal,roccati2022non}
\begin{equation}\label{eq:master_equation_approx}
\begin{aligned}
\frac{\partial \rho} {\partial t} = -i [H_{\rm NH}, \rho] + 2\gamma z \rho z^{\dagger}
\end{aligned}
\end{equation}
where the second term corresponds to quantum jumps and can be dropped off in the semi-classical limit~\cite{carmichael1993quantum,minganti2019quantum}. Meanwhile, the effective non-Hermitian Hamiltonian is
\begin{equation}\label{eq:NH_Hamiltonian}
\begin{aligned}
H_{\rm NH} & \equiv H_{AB} - i \gamma z^{\dagger} z \\
& = -i \gamma ( S^{+}_A S^{-}_A + S^{+}_B S^{-}_B + 2 e^{i k_z d} S^{-}_A S^{+}_B ).\\
%\gamma & = \frac{2g_A g_B}{\dp}
\end{aligned}
\end{equation}
Notably, apart from the decay into the phonon mode (first two term), $H_{\rm NH}$ describes a non-reciprocal interaction $S_A^{-} S_B^{+}$ mediated by the $(+ k_z, + L)$ phonon - Spin A can give its spin angular momentum to spin B, but not vice versa (upper panel of Figure~\ref{fig:non-reciproal}).  This is consistent with our intuitive arguments above. Note that this is an off-diagonal non-Hermitian interaction, instead of the more common diagonal non-Hermitian interaction induced by e.g., gain and loss~\cite{wu2019observation}.

However, one must also consider the effect of the $(-k_z, +L)$ phonon mode, which carries a PAM of $+1$, but is left-propagating (lower panel of Figure~\ref{fig:non-reciproal}). Obviously, this phonon tends to transfers a $+1$ angular momentum from spin B to spin A. Using the arguments above, one can show that  $(-k_z, +L)$ leads to a $-2i \gamma'  e^{-i k_z d} S^{+}_A S^{-}_B$ term in the total spin-spin interaction~\cite{carmichael1993quantum}. Here the coupling strength is $\gamma' = \frac{2g'^2}{\Delta'}$, where $g'$ is the spin-phonon coupling strength with the  $(-k_z, +L)$ phonon mode, while $\Delta'$ is the frequency detuning. One can see that the non-reciprocal interactions induced by the two counter-propagating phonons $(\pm k_z, +L)$ tends to counter-balance each other. The interaction becomes fully reciprocal when $\gamma = \gamma'$, which is the case for non-chiral material, in which $(\pm k_z, +L)$ are degenerate. Contrarily, in chiral materials, one would have $g'\neq g$ and $\Delta' \neq \Delta$. While $g'$ and $g$ are typically comparable in magnitude, it is possible to have $\Delta' \gg \Delta$ by judiciously choosing the spin and phonon frequencies. This would result in a highly non-reciprocal interaction $\gamma' \ll \gamma$ between spins A and B, as we will show later.

Besides, the other two modes $(+k_z, -L)$ and $(-k_z, -L)$ can contribute to spin-spin interactions as well. However, as mentioned earlier, they involve counter-rotating terms such as $S_A^{+} a^{\dagger}$, leading to extremely large detuning [cf. Eq.~(\ref{eq:spin_phonon_total_Hamitonian})]. The resultant spin-spin coupling strength is thus small and negligible~\cite{winkler2003quasi}. The spin-spin interactions mediated by all four phonon modes $(\pm k_z, \pm L)$  are summarized in Table~\ref{tab:four_modes}.

\textbf{Experimental Realizations.} In this section, we will discuss possible experimental realization of the non-Hermitian spin-spin interaction mediated by chiral phonons and estimate its strength. We will first consider electron spins, whose frequencies are typically on GHz order.

We assume the two electron spins are embedded in a mechanical resonator with dimension $(l,w,h)$, which is made of chiral materials such as $\alpha$-$\rm SiO_2$. With $l \gg w, h$, the mechanical modes supported by resonator should have wave vectors $k_z = \frac{n\pi}{l}$, and the corresponding eigen-frequency is $\omega_{\pm} = v_{\pm} k$, where we use $\pm$ to denote modes with $\pm k_z$. Note that this is a simplified picture and more accurate properties of the mechanical modes can be obtained by solving the corresponding wave equations.  For electron spin, we will fix $l=1~\rm \mu m$ and $w=h=0.1~\rm \mu m$, leading to a volume of $V = 0.01~\rm \mu m^3$~\cite{bennett2013phonon,fung2023programmable,rabl2010quantum,li2020enhancing}. The fundamental mode ($n=1$) frequency is $\omega_{\pm} \sim 1~\rm GHz$ in $\alpha$-$\rm SiO_2$. Other modes with $n>1$ can be ignored since they are far off-resonance with the electron spins. The zero-point field of the TA modes can be obtained from $u^{\rm zpf} = \sqrt{\frac{\hbar \omega}{2 \rho v^2 V}}$, where $\rho$ is the density of $\alpha$-$\rm SiO_2$. This leads to $g = \Xi^S u^{\rm zpf} \sim 1~\rm kHz$. Here we do not distinguish $g$ for the $\pm k_z$ modes since they are comparable. For the $(+k_z, +L)$ mode, we can tune the spin frequency so that the detuning is $\Delta = 10~\rm kHz$, leading to a spin-spin interaction of $\gamma \sim 0.1~\rm kHz$. Potentially $\gamma \gtrsim 1~\rm kHz$ can be realized by using a resonator with smaller dimension, or by using defects with larger $\Xi^S$. Since the decoherence time  $T_2$ of spins in $\rm SiO_2$ can potentially reach several milliseconds~\cite{kanai2022generalized}, we expect $\gamma\sim [0.1-1]~\rm kHz$ would lead to observable effects.

On the other hand, due to the $20~\%$ difference in $\omega_{\pm}$ (Figure~\ref{fig:chiral_phonon}), the frequency detuning for the $(-k_z, +L)$ mode would be as large as $\Delta' \sim 0.1~\rm GHz$, leading to an extremely small $\gamma' \sim 0.01~\rm Hz$. The spin-spin interaction is thus strongly non-reciprocal since one has $\gamma' \ll \gamma$.

For nuclear spins, the dimension of the resonator needs to be on the order of $l\sim 1~\rm mm$ to support mechanical waves with $\rm MHz$ frequency, to be comparable with nuclear spin frequencies. This leads to a nuclear spin-phonon coupling strength $g \sim 10^{-4}~\rm Hz$, which can be too weak to induce any observable effects. To improve $g$, one may externally drive the mechanical waves in the resonator~\cite{schuetz2017high}. For example, if the strain field of the driven mechanical wave is $u = 10^{-4}$, then the nuclear spin-phonon coupling strength would be $g\sim \Xi^I u \approx 100~\rm Hz$, which is a relatively strong interaction for nuclear spins. Note that an externally driven mechanical wave may lead to side effects such as faster spin decoherence. As can be seen from Eq.~(\ref{eq:NH_Hamiltonian}), both the non-reciprocal spin-spin interaction and spin decoherence scales as $\gamma \propto u^2$.  However, it can still be beneficial there are other decoherence $\gamma_{0}$ due to e.g., charge noise. With a driven $u$ field, one may achieve $\gamma \gg \gamma_0$, so that the non-Hermitian dynamics would be more pronounced. Here it is worth mentioning that it is not detrimental to have multiple phonons on the desired phonon mode ($n=1$ mode here). This is because while they lead to faster spin relaxation,  they also contribute to stronger spin-spin interaction $\gamma$ . On the other hand, it would be advantageous to suppress the number of phonons on other modes (i.e., $n\neq1$).

%If $\gamma \ll \gamma_0$, then the non-Hermitian dynamics cannot be probed. On the other hand. if $\gamma$ is to be compete with other decoherence channels due to e.g., charge noise.

\textbf{Discussions.} The non-Hermitian spin-spin interactions mediated by chiral phonons can be extended to multiple spins. For a chain of spins embedded in chiral materials, only spins on the left can give positive angular momentum to spins on the right, but not vice versa. This is a prototypical cascaded quantum systems, which can be used for e.g.,  entangled state preparation~\cite{stannigel2012driven}. The phase of the spin-spin interaction is dependent on the phase factor $e^{ikd}$, leading to a highly tunable Hamiltonian, which is a good candidate for studying non-Hermitian many-body physics. Furthermore, since the transport of spin excitation is uni-directional, the recently proposed exponentially enhanced non-Hermitian cooling~\cite{xu2024exponentially} may also be demonstrated.

In chiral materials, circularly polarized light with $\pm 1$ helicity also have different properties such as refractive index, which can potentially leads to non-Hermitian spin-spin interaction as well. However, as we demonstrate in the SI Section B, the photon-mediated spin-spin interaction is much weaker than the phonon-mediated interaction. This is mainly because the large photon group velocity, which leads to long wavelength and large resonator mode volume at a specific frequency. Additionally, while electrons in chiral materials also exhibit spin-momentum locking, they cannot lead to non-Hermitian interactions between localized spins. This limitation arises from the Fermionic nature of electrons - they cannot be simply created or annihilated like phonons. Instead, when electrons undergo spin flip-flop interactions with localized spins, they are scattered between spin-up and spin-down states. As a result, these two spin states are coupled together, thereby preventing any non-Hermiticity that might otherwise emerge from the chirality of the material (details in  SI Section C).

In summary, we proposed non-Hermitian spin-spin interactions mediated by chiral phonons. The coupling between phonon momentum and angular momentum results in selective and non-reciprocal spin-spin interactions. The long-range nature of acoustic phonons enables the realization of long-sought non-Hermitian interactions among multiple spins. This advancement opens up new possibilities for applications in many fields, such as non-Hermitian many-body physics.

%A simple but generic answer is: the phonon must have different properties and thus behave differently when (virtually) traveling from spin A to B compared to travelling from B to A. The spin-phonon interaction relies on the angular momentum of the phonons, while the direction of travel is encoded in the momentum of the phonons. Therefore, to mediate spin-spin interaction in a non-Hermitian fashion, the angular momentum must be entangled with the momentum of the phonon. This is similar to the spin-Hall effect, which requires spin-orbit interaction to lock the momentum and the spin, so that electrons with certain spin polarization can travel in a preferred direction\cite{sinova2015spin}, lead to the spin Hall current. However, conventional phonon modes are linearly polarized, which is the equal superposition of states with opposite angular momentum.  Hence, the is no

% \begin{equation}\label{eq:HAB_sigma_with_phi}
% \begin{aligned}
% \frac{\partial }{\partial t} \langle \sigma^{-}_A \rangle & = -\gamma \langle \sigma^{-}_A \rangle \\
% \frac{\partial }{\partial t} \langle \sigma^{-}_B \rangle & = -\gamma \langle \sigma^{-}_B \rangle - 2 \gamma  e^{-ikd}\langle \sigma^{-}_A \rangle
% \end{aligned}
% \end{equation}

%\bibliographystyle{apsrev4-2}
\bibliography{main}% Produces the bibliography via BibTeX.

% ****** End of file apssamp.tex ******

\end{document}